\documentstyle[preprint,aps,epsf]{revtex}
\begin{document}
\draft
\title{Interfacial Structural Changes and Singularities in
Non-Planar Geometries}

\author{C.\ Rasc\'{o}n$^{1,2}$, A.O.\ Parry$^1$}
\address{$^1$Mathematics Department, Imperial College\\
180 Queen's Gate, London SW7 2BZ, United Kingdom}
\address{$^2$Departamento de F\'{\i}sica Te\'{o}rica de la Materia
Condensada,\\ Universidad Aut\'{o}noma, E-28049 Madrid, Spain}
\date{\today}
\maketitle

\begin{abstract}
We consider phase coexistence and criticality in a thin-film Ising magnet
with opposing surface fields and non-planar (corrugated) walls.
We show that the loss of translational invariance has a strong and unexpected
non-linear influence on the interface structure and phase diagram.
We identify 4 non-thermodynamic singularities where there is a
qualitative change in the interface shape. In addition,
we establish that at the finite-size critical point, the singularity
in the interface shape is characterized by two distint critical
exponents in contrast to the planar case (which is characterised
by one). Similar effects should be observed for prewetting at a
corrugated substrate. Analogy is made with the behaviour of a
non-linear forced oscillator showing chaotic dynamics.
\end{abstract}
\pacs{ PACS numbers: 68.45.Gd, 68.10.-m, 68.35.Rh}

There are a number of well studied examples of fluid interfacial
phenomena for planar systems in which surface phases with distinct
adsorptions co-exist along a line of first-order phase transitions
which terminates at a surface critical point. Examples include the
pre-wetting transition associated with first-order wetting
\cite{Dietrich} and also interfacial localization in thin-film magnets
(with opposite surface fields) associated with confinement effects at critical
wetting \cite{PE}. In both cases, the difference in adsorption between
the two phases vanishes continuously as the critical point, signifying
the end of two-phase coexistence, is approached. This second order phase
transition is characterized by the critical exponents belonging to the
two dimensional Ising universality class (for three dimensional bulk
systems) since the adsorption difference acts as a scalar order
parameter \cite{BLF,NF}. In this letter, we describe a wealth of
new interfacial structural changes and singularities which emerge
when the analogous phenomena are considered in slightly non-planar
geometries and which are intimately associated with non-linear behaviour.
In addition to a shift in the finite-size (FS) critical point (compared to the
planar confined system), the shape of the non-planar interface
undergoes a number of structural changes as we move along and beyond
the line of coexistence. This behaviour has no counterpart in the planar
geometry, and has not been reported previously.
Moreover, as the shifted surface critical
point is approached, the function describing the shape of the non-planar
interface shows non-analyticities which are characterised by two critical
exponents. Whilst one of these appears to be identical with the usual
critical exponent describing the singularity in the total (or average)
adsorption, the general identification of the second is a more difficult
problem, although scaling arguments (consistent with our explicit results)
suggest that its value is related to the energy density.
 Our predictions are based on a detailed numerical analysis
of a simple mean-field (MF) model of interfacial behaviour which we
believe is qualitatively correct beyond MF approximation (in three
dimensions). These are rather dramatic effects emanating from the
introduction of a slight non-planar perturbation to the interface
can be viewed profitably by making analogy with the classical
mechanics of an extremely sensitive non-linear dynamical system
exhibiting chaotic behaviour \cite{Chaos}. As we will see, the
interface behaviour may be elegantly portrayed as the temperature
evolution of a phase plane plot, similar to that employed in
dynamical systems, allowing us to distinguish different qualitative
types of interface shape separated by non-thermodynamic singularities.

To begin, we recall the relevant properties and phase diagram of the
planar system prior to a discussion of the non-planar generalization.
The transition that we concentrate on occurs in a thin-film magnet
with opposing surface fields but the phenomena are generic to other
situations such as prewetting at a planar wall. Consider then an
Ising-like thin film magnet of width $L_z$ and infinite transverse
area in zero bulk field and below the bulk critical temperature
$T^{BULK}_{c}$ with surface fields $h_1$ and $h_2\!=\!-\!h_1$ acing on the
spins in the $z\!=\!0$ and $z\!=\!L_z$ planes respectively. We further suppose
(through a judicial choice of surface enhancement \cite{NF}) that in the
semi-infinite limit $L_z\rightarrow\infty$ each surface undergoes a
critical wetting transition at temperature $T_w$. For such a system,
MF \cite{PE} and simulation studies \cite{BLF} show that the finite
size phase diagram is dominated by wetting effects which are able to suppress
bulk-like coexistence over a large temperature regime. At sufficiently
low temperatures $T\!<\!T_c(L_z)$, with the finite-size critical
temperature satisfying $T_c(L_z)\!<\!T_w$, phase coexistence is possible
between phases corresponding to an interface being bound to either wall.
As the temperature increases, the interface position moves continuously
to the middle of the system and for $T\!>\!T_c(L_z)$ only one phase is
possible. Thus, in the temperature window $T^{BULK}_{c}\!>\!T\!>\!T_w$,
the FS effects suppress bulk-like phase coexistence for all
$L_z$. This temperature range is also characterised by a near soft-mode
phase since the transverse correlation length $\xi_\parallel$ is
extremely large due to capillary-wave like excitations. These features
can be most easily understood using a simple effective interfacial
Hamiltonian model \cite{PE}:
\begin{equation}
\label{one}
H[\ell]=\int\!d{\bf r} \left[{\Sigma\over 2}(\nabla\ell)^2+W(\ell;L_z)\right]
\end{equation}
where $\ell ({\bf r})$ is the collective co-ordinate describing the
interface position at vector displacement ${\bf r}\!\!=\!\!(x,y)$
along the wall and $\Sigma$ is the stiffness coefficient of the
up-spin-down-spin interface. The total finite-size binding potential
$W(\ell;L_z)$ acting on the interface (whose minima determine the
MF location/s of the interface) is the sum of the two contributions from
each wall:
\begin{equation}
\label{two}
W(\ell;L_z)=W_{\infty}(\ell)+W_{\infty}(L_z-\ell)
\end{equation}
where $W_{\infty}(\ell)$ is the appropriate semi-infinite binding
potential for the ranges of forces in the model. For systems with
short ranged forces this is usually specified as \cite{Dietrich}
\begin{equation}
\label{three}
W(\ell)=a_{o}(T-T_w)\;e^{-\kappa\ell}+b_{o}\;e^{-2\kappa\ell}\; ;
\hspace{1cm}\ell>0
\end{equation}
with $a_{o},b_{o}$ positive constants and $\kappa$ being the inverse
bulk correlation length. For $T\!<\!T_c(L_z)$, with
$T_c(L_z)\!=\!T_w-4(b_{o}/a_{o})e^{-\kappa L_z/2}$ in MF approximation,
the total potential $W(\ell;L_z)$ has a double well
structure with two equal minima at $\ell_{\pi}<L_z/2$ and
$\ell_{\pi}^{*}=L_z-\ell_{\pi}$. As $T\!\rightarrow\!T_c(L_z)^{-}$,
the adsorption difference $\Delta\Gamma=4m_{o}(L_z/2-\ell_{\pi})$
(with $m_{o}$ the bulk magnetization) vanishes like
$\Delta\Gamma\sim(T_c(L_z)-T_c)^{1/2}$, corresponding to a
standard order-disorder transition. For $T\!>\!T_c(L_z)$,
the potential $W(\ell;L_z)$ has only one minimum at $\ell_{\pi}\!=\!L_z/2$
and the correlation length $\xi_\parallel\sim e^{\kappa L_z/4}$.
Interestingly, most of these quantitative MF predictions are
confirmed by extensive Monte Carlo simulation
studies which established
that the true asymptotic critical regime where we can expect
Onsager-like behaviour $\Delta\Gamma\sim(T_c(L_z)-T_c)^{1/8}$
is extremely small \cite{BLF}. All these facts support MF theory as
an excellent quantitative description of the thin film system.

We now wish to consider the MF phase diagram for the analogous
phase transition in a slightly non-planar geometry.
We will take as our starting point the simplest possible
phenomenological model of this system which generalises (\ref{one})
and suppose that the configuration energy is specified by
\begin{equation}
\label{four}
H[\ell;z_w^{(1)},z_w^{(2)}]=\int\!d{\bf r} \left[{\Sigma\over 2}
(\nabla\ell)^2+W(\ell;L_z,z_w^{(1)},z_w^{(2)})\right],
\end{equation}
where $z_w^{(1)}({\bf r})$ and $z_w^{(2)}({\bf r})$ describe the
(small) deviations of the walls
near the $z=0$ and $z=L_z$ respectively and $W(\ell;L_z,z_w^{(1)},z_w^{(2)})=
W_{\infty}(\ell-z_w^{(1)})+W_{\infty}(L_z+z_w^{(2)}-\ell)$.
Whilst the model could certainly be improved by including further
coupling terms involving $\nabla\ell\cdot\nabla z_w$ with associated
position dependent (stiffness) coefficients, we do not expect these
to make any significant difference to the interfacial behaviour
described here \cite{RP}. In any case, even with the further assumption of
corrugated walls such that $z_w^{(1)}$ and $z_w^{(2)}$ only depend on a
single-coordinate ($x$ say), the interfacial behaviour generated is
sufficiently complex to warrant attention within the simple model above.
Writing $z_w^{(1)}(x)=a\sqrt{2}\sin({\sl q}x)$, we have considered
the geometry for which $z_w^{(1)}\!=\!z_w^{(2)}$
although, of course, many other choices are possible \cite{RP}. The
r.m.s.\ width $a$ and wavelength $L_x\!=\!2\pi/{\sl q}$ of the wall
corrugation are assumed to be small and large respectively in comparison with
the bulk correlation length. With these assumptions, the wetting
transition remains second order and located at $T_w$ in the semi-infinite
limit \cite{PSF}.

The equilibrium non-planar
interfacial profile/s $\ell_{\nu}(x)$ satisfies the Euler-Lagrange
equation
\begin{equation}
\Sigma\;\ddot{\ell}_{\nu}(x)=W_{\infty}'(\ell_{\nu}-z_w^{(1)})
-W_{\infty}'(L_z+z_w^{(2)}-\ell_{\nu})
\label{five}
\end{equation}
where dot and prime signify differentiation w.r.t.\ $x$ and
argument respectively. Periodic boundary conditions are imposed
after a large multiple of wavelengths $L_x$. Two preliminary
remarks are as follows: firstly, the Euler-Lagrange equation is
inversion symmetric so that if ${\ell}_{\nu}(x)$ is a solution,
${\ell}_{\nu}^{*}(x)\!=\!L_z-{\ell}_{\nu}(x\!+\!\pi/{\sl q})$
is also a solution with the same free energy and is distinct from
${\ell}_{\nu}(x)$ in the two-phase regime. Secondly, we have established
numerically that the stable phases all have the same wavelength as
the wall corrugation $L_x$. However, this is not the case for the
metastable states \cite{RP}. Finally, we note that an elegant
description of the interfacial shape is afforded by a reduce phase plane
plot $\ell/\sqrt{2}a$ {\it vs.} $\dot{\ell}/\sqrt{2}a{\sl q}$ and helps
distinguish different types of structural regimes. A section of the
equilibrium phase diagram, with suitable reduced units \cite{units},
is shown in Fig.\ $1$ and shows a critical
line (corresponding to an order-disorder transition) and four
non-thermodynamic singularities where there is a qualitative change of
interfacial structure. In this way, we are able to distinguish five different
interfacial types (see Fig.\ $2$).

Phase coexistence and order is most easily revealed through the mean
interfacial height
\begin{equation}
\label{seven}
\ell_{o}\equiv {1\over{L_x}}\int_{0}^{L_x}\!\!dx\;\,\ell_{\nu}(x)
\end{equation}
which is single valued ($\ell_{o}\!=\!L_z/2$) in the disordered regime
above the critical temperature $T_c(L_z,a,{\sl q})$, but is double
valued (with $\ell_{o}^{*}\!=\!L_z-\ell_{o}$) in the order regime,
analogous to $\ell_{\pi}$ and $\ell_{\pi}^{*}$ for the planar system.
Our numerics indicate that the singularity in $\ell_{o}$ is of the expected
type:
\begin{equation}
\label{eight}
{{L_z}\over 2}-\ell_{o}\simeq
\left\{ \begin{array}{ll}
t^{1\over 2} & \hspace{1cm} \hbox{if}\,\,\,\, t>0 \\ \\
0 & \hspace{1cm} \hbox{if}\,\,\,\, t<0
\end{array} \right.
\end{equation}
where we have introduced the scaled temperature variable
$t\!\equiv\!(T_c(L_z,a,{\sl q})-T)/T$. In addition to the mean interface
height, however, the shape function shows a number of qualitative changes
with temperature. At very low temperatures, the interface is closely
bound to one of the walls and follows the corrugation (See Fig. $2$(A)).
Over one period $L_x$, the graph
$\ell_{\nu}(x)$ has one maximum and one minimum which are in phase with
the wall function $z_w^{(1)}(x)$. For this case, the phase plane plot
is a simple loop. Nevertheless, notice that its form is not precisely
circular, indicating that non-linear effects are important even when
the interface is close to the wall. On increasing the temperature, the
interface smoothly deforms and shows a number of non-thermodynamic
singularities where the minima and maxima of the graph undergo a series
of {\it bifurcations}. These reveal themselves as the
appearance/disappearance of loops in the phase plane portrait as illustrated
in Fig.\ $2$(B) which also shows the locus of the maxima/minima
with temperature (Fig.\ $2$(C)). Corresponding profiles are shown in
Fig.\ $2$(A).
Two counter intuitive features are worth emphasising here. Firstly,
there are two regimes, {\it II} and {\it IV},
where the interface shape has two and three
maxima per wavelength of the wall corrugation. Secondly, in the
vicinity of the order-disorder transition, regime {\it III},
the interface shape is
similar to the wall ({\it i.e.} there is only one max/min pair per period)
but is {\it out of phase} with it. Finally, at high temperatures
above the two super critical non-thermodynamic singularities, the
interface shape returns to that of a simple sinusoidal-like function in phase
with the wall and the phase portrait is basically a circle of radius
$(1+{\sl q}^{2}\xi_\parallel^{2})^{-1}$ centred at $L_z/2$.

Next, we focus on the singularity in the shape profile at the
order-disorder transition. We have established that the stable
phase/s can be represented by a Fourier series
\begin{eqnarray}
\label{nine}
\ell_{\nu}(x)\!=\!\ell_{o}+\sigma_{1}\sin({\sl q}x)+\hspace{1.5cm}\\
\sum_{k=1}^{\infty}\left\{\sigma_{2k+1}\sin\left((2k+1){\sl q}x\right)
+\gamma_{2k}\cos(2k{\sl q}x)\nonumber
\right\}
\end{eqnarray}
throughout the phase diagram. In this expression $\ell_{o}$ is
the mean interface position (given by Eq.\ (\ref{seven})), whilst
the second term is the harmonic response to the wall corrugation.
The final term represents the higher order harmonic excitations arising
from the non-linearity of the Euler-Lagrange equation and are
responsible for the complicated evolution of the interface structure with
temperature. We stress that, without this term ({\it i.e.\ }
just considering linear response), the phase plane portrait would be
simply circular.
Note that there are not even sine terms and no odd cosine terms.
The temperature dependence of the two sets of coefficients
$\{\sigma_{2k-1}\}$ and $\{\gamma_{2k}\}$ is extremely
involved but near $T_c(L_z,a,{\sl q})$ only two types of singularity
are observed in our numerical analysis (See Fig.\ $3$).
The coefficients $\{\gamma_{2k}\}$ all vanish above
$T_c(L_z,a,{\sl q})$ and behave
precisely as the mean order-parameter ${{L_z}\over 2}-\ell_{o}$,
{\it i.e.} they are characterised by the usual MF order-parameter
critical exponent $\beta\!=\!1/2$. In contrast, the terms
$\{\sigma_{2k-1}\}$ all have a cusp-like singularity
\begin{equation}
\label{ten}
\sigma_{2k-1}-\sigma_{2k-1}^{c}\simeq\;\,|t|^{\theta}\,\, ;
\hspace{1cm}t\!\rightarrow\! 0^{\pm}
\end{equation}
where $\sigma_{2k-1}^{c}$ is the value at criticality and the critical
exponent $\theta\!=\!1$. There is no analogy of this singularity in
the planar system. Furthermore, whilst it is natural to identify
the cosine term singularities with the order-parameter exponent $\beta$
of the $d-1$ dimensional bulk universality class ($\beta\!=\!1/2$ in
MF, $\beta\!=\!1/8$ beyond MF for three dimensional thin films), a 
similar identification for $\theta$ is not so obvious.
Nevertheless, we have constructed scaling arguments which
suggest that $\theta\!=\!1\!-\!\alpha$, where $\alpha$ is the
specific heat critical exponent, consistent with our
numerical results \cite{RP}. Similarly, we have also established that,
for fixed $L_z$, the critical line is consistent with the scaling law
\begin{equation}
\label{eleven}
T_c(L_z,a,{\sl q})-T_c(L_z,0,0)\simeq\;a^{2}\;\Lambda({a\over{\sl q}})
\end{equation}
where $\Lambda$ is an appropriate scaling function.
This behaviour can be understood using finite-size scaling ideas with MF
critical exponents and indicates that the effective width of the system
is reduced by corrugation \cite{RP}.

To finish our article, we make some pertinent remarks. Firstly,
the interfacial structural changes reported here are not peculiar to
short-ranged forces with the exponential binding potential Eq.\
(\ref{three}), and also emerge if long-ranged forces are considered
instead \cite{RP}. Also, we emphasise
that in previous studies of
the effect of roughness on wetting transitions most authors have considered
binding potentials with a single minimum which do not exhibit the same
subtle non-linear behaviour discussed here \cite{PSF,Others}.
Next, we note that on making a change of variable
$\eta(x)\!\equiv\!\ell(x)\!-\!L_z\!-\!z_w^{(1)}$ and expanding to
appropriate (cubic) order, the Euler-Lagrange equation can be written
\begin{equation}
\label{twelve}
\Sigma\;\ddot{\eta}=-\tilde{t}\eta+\tilde{u}\eta^{3}+
a{\sl q}^2\sin({\sl q}x)
\end{equation}
where $\tilde{t}\propto(T_c(L_z)-T)$ and $\tilde{u}$ is
positive in the region of interest. This is essentially equivalent
to the Duffing equation of a soft-polynomial oscillator
(without a damping term) which is known to yield extremely rich
(including chaotic) dynamics \cite{Chaos}. In this context, the
non-thermodynamic singularities described above are analogous to
the harmonic excitations of the non-linear oscillator (however,
this analogy does not shed any light on the nature of the
singularities near the order-disorder transition and their
identification beyond MF).

In summary, we have shown from a simple MF model of interfacial behaviour
in a slightly non-planar geometry that new types of structural phase changes
and additional critical singularities can emerge which are intimately
related to non-linear phenomena. Similar behaviour is also expected
for pre-wetting at a non-planar substrate \cite{RP}. Also of interest is the structure
of metastable states in this system which we do not discuss here \cite{RP}.
We believe that future studies of improved models which include thermal
fluctuations and different type of non-planarity will also uncover new
structural and fluctuation related behaviour.

One of us (C.R.) is grateful to E.\ Velasco for friendly discussions
and acknowledges economical support from {\it La Caixa} and
The British Council.

\begin{figure}
\label{first}
\vspace*{3cm}
\caption{Phase diagram for $L_z\!=\!10$ and $q\!=\!2\pi/10$
in reduced units \protect\cite{units}. The solid line separates the ordered
and the disordered phases. The dashed lines show the
location of the non-thermodynamic singularities and divide the
phase diagram into 5 regions.}
\end{figure}

\begin{figure}[p]
\label{second}
\vspace*{2cm}
\caption{Behaviour of the system for $L_z\!=\!10$, $a\!=\!1.5$
and $q\!=\!2\pi/10$ showing the
shape of the interface in the different regimes (A) and their
corresponding phase portraits $\dot{\ell}\;\hbox{\it vs.\ }\ell$ (B).
The circle represents the point $x\!=\!0$. For clarity, scales related
to $\ell$ are ommited but can be checked in Fig.\ $3$.
The loci of the interface minima and maxima are represented as a
funcion of the temperature (C). The FS critical temperature
$T_c(L_z,a,{\sl q})\!\approx\!0.845$ is represented by a thin line
and is within regime $III$.}
\end{figure}

\begin{figure}[p]
\label{third}
\vspace*{2cm}
\caption{Behaviour of the coefficients $L_z/2-\ell_{o}$, $\sigma_{1}$,
$\gamma_{2}$ and $\sigma_{3}$ of Eq.\ \protect\ref{nine} near
$T_c(L_z,a,{\sl q})$ for $L_z\!=\!10$, $a\!=\!1.5$ $q\!=\!2\pi/10$.
They are multiplied by $1$, $10L_x$, $10^{2}L_x$ and $10^{3}L_x$,
respectively.}
\end{figure}


\begin{references}
\bibitem{units} Throughout the paper, lengths are measured in (dimensionless)
units of the bulk correlation length and temperature in units of $T_w$
(the wetting temperature of the semi-infinite system).
\end{references}
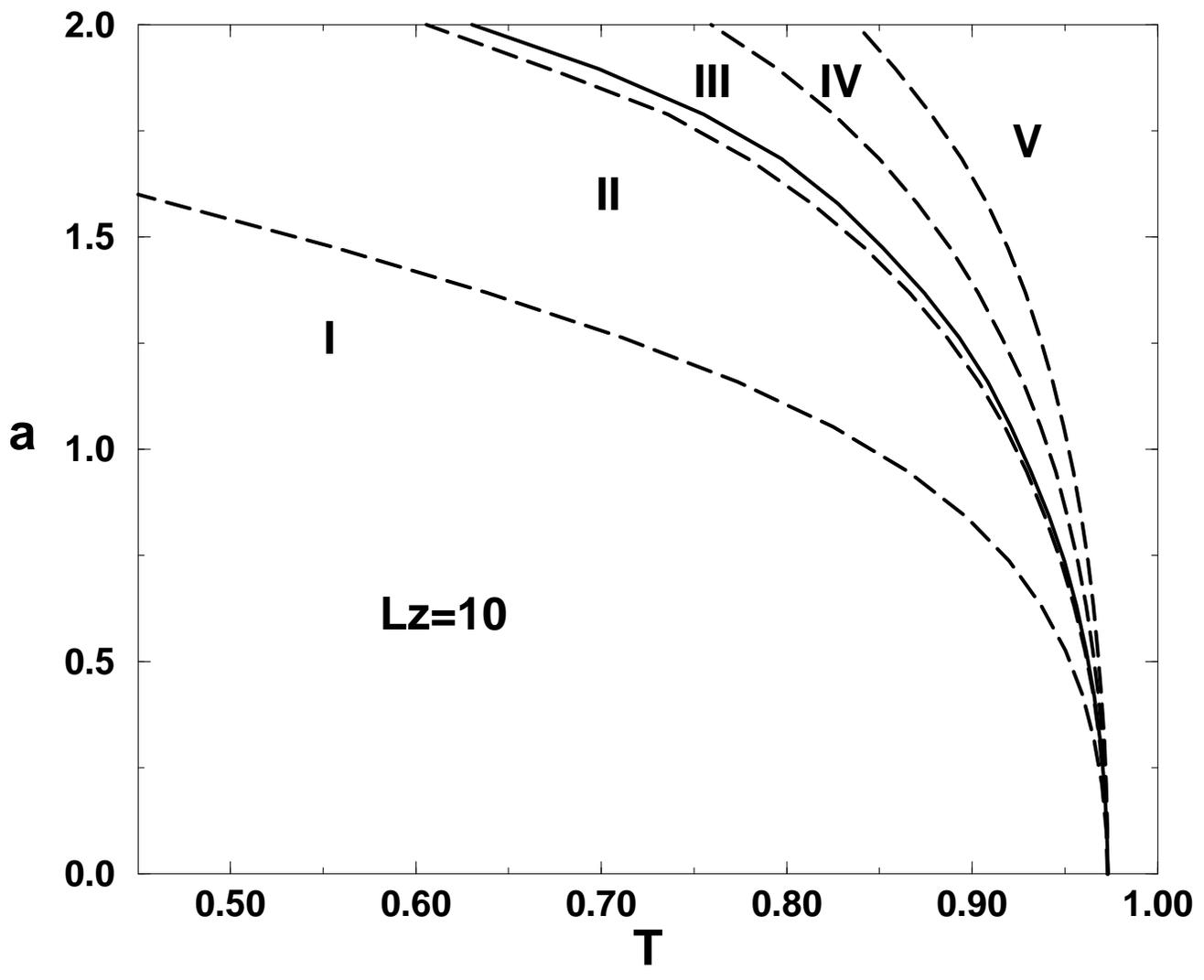
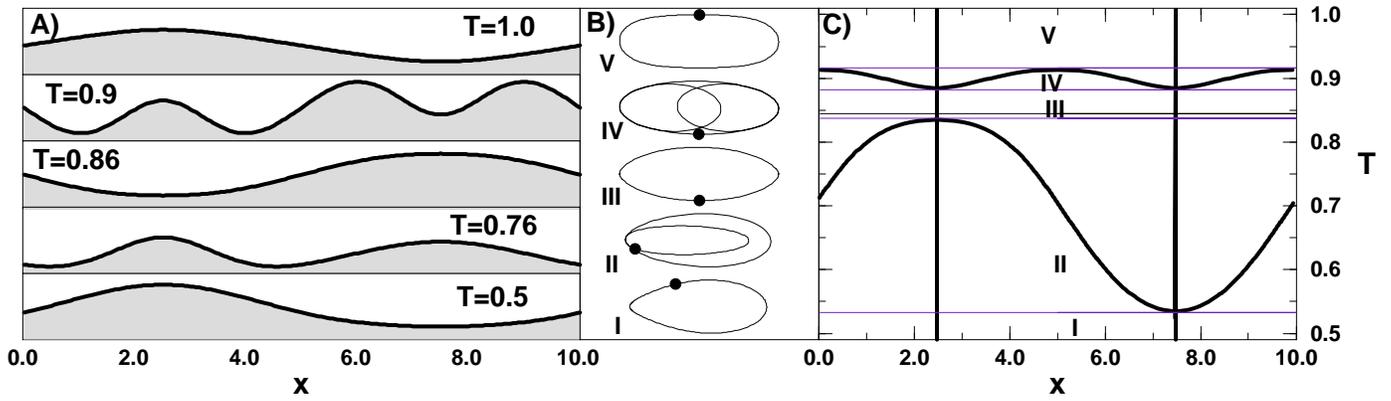
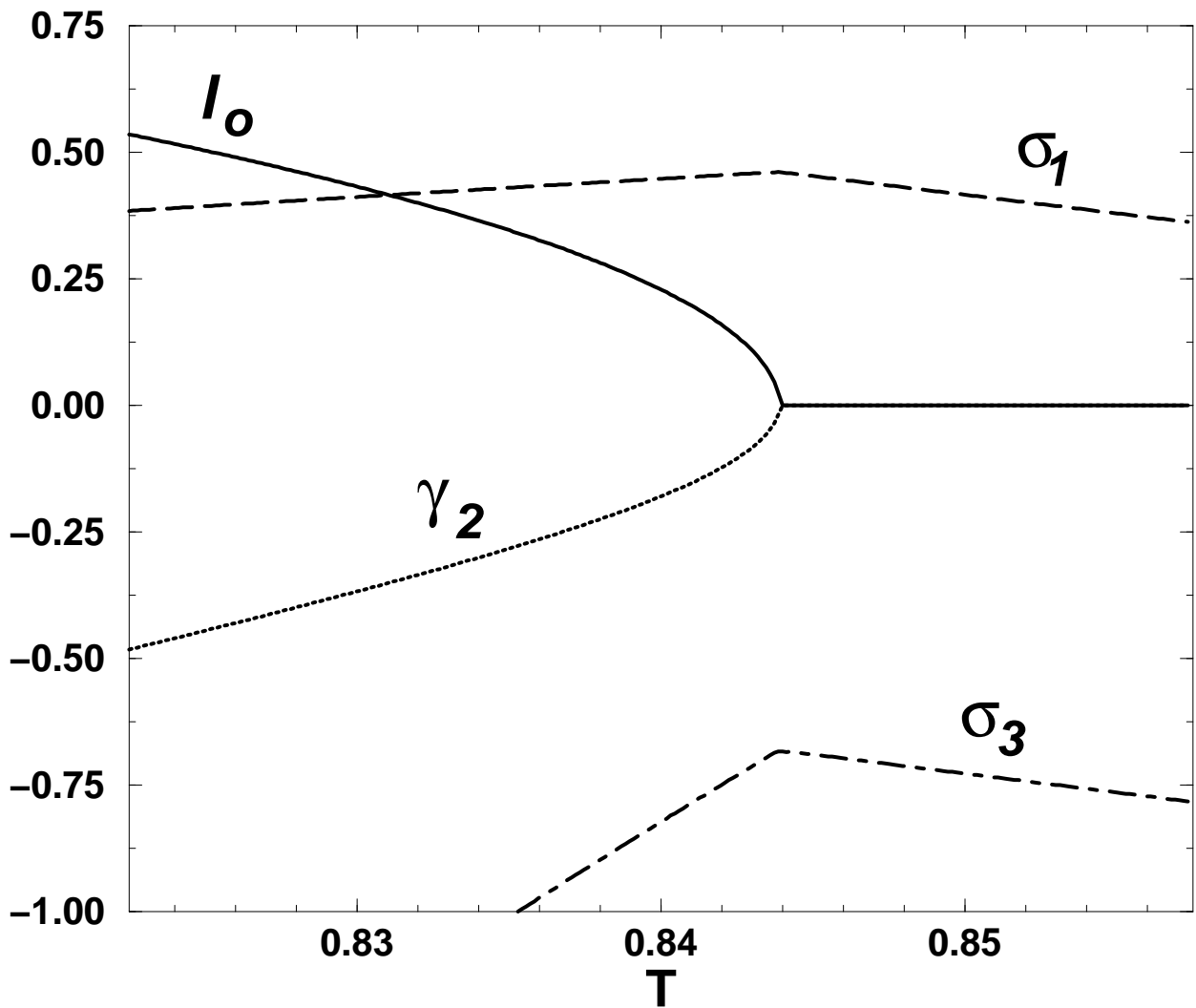
\end{document}